\documentclass[aps,pra,groupedaddress,showpacs,preprint]{revtex4-1}
\usepackage{graphicx}

\begin{document}

\title{Potential-splitting approach applied to the Temkin-Poet model
for electron scattering off the hydrogen atom and the helium ion}

\author{E.~Yarevsky$^1$}
\email[E-mail: ]{yarevsky@gmail.com}
\author{S.~L.~Yakovlev$^1$}
\email[E-mail: ]{s.yakovlev@spbu.ru}
\author{{\AA}.~Larson$^2$}
\email[E-mail: ]{aasal@fysik.su.se}
\author{N.~Elander$^2$}
\email[E-mail: ]{elander@fysik.su.se}

\affiliation{$^1$ Department of Computational
Physics, St Petersburg State University, 198504 St Petersburg, Russia }
\affiliation{$^2$ Chemical Physics Division, Department of Physics, AlbaNova University Centre, Stockholm University, 106 91 Stockholm, Sweden, EU}

\begin{abstract}
The study of scattering processes in few body systems is a difficult problem especially if long range interactions are involved.
In order to solve such problems, we develop here a potential-splitting approach for three-body systems.
This approach is based on splitting the reaction potential into a finite range core part and a long range tail part.
The solution to the Schr\"odinger equation for the long range tail Hamiltonian is found analytically, and used as an incoming wave in the three body scattering problem.
This reformulation of the scattering problem makes it suitable for treatment by the exterior complex scaling technique in the sense that the problem after the complex dilation is reduced to a boundary value problem with zero boundary conditions.
We illustrate the method with calculations on the electron scattering off the hydrogen atom and the positive helium ion in the frame of the Temkin-Poet model.
\end{abstract}


\maketitle

\section{Introduction}

The Coulomb force is the basic interaction mechanism in atomic and molecular physics.
However, solving the Coulomb scattering problem even for few particles is a difficult task from both theoretical and computational points of view.
The reason for this complexity is concealed in the long-range character of the Coulomb interaction.
Here, the asymptotic boundary conditions for the wave function at large distances are complicated for the few-body scattering problem~\cite{FadMerk}.
An analytic solution for the Coulomb problem does not exist if three or more particles are involved in the scattering process.
For such systems, methods which allow solving the problem without explicit use of the asymptotic form of the wave function are of great importance.

One such approach is the complex scaling method.
Originally, this technique was based on the uniform dilation of coordinates~\cite{NuttCo69,BalCom71} and could only be applied to problems involving either finite range or exponentially decreasing potentials.
This method was subsequently modified in such a way that some longer range potentials (not, however, a Coulomb potential in the asymptotic region) could be studied~\cite{RescBaerByrMcC_long1997}.
Essentially, the modification consisted in replacing the potential $V(r)$ in the problem by a finite cut potential $V_R(r)=V(r)$, for $r<R$ and $V_R(r)=0$ for $r\geq R$ with the believe that results for $V_R$ as $R\to \infty$ will recover the solution of the original problem.
As $V_R(r)$ is not an analytic function, the exterior complex scaling (ECS) method was employed.
This approach has been successfully used for three-body electron-hydrogen scattering calculations~\cite{McCRescByr_ioniz1997}.

We emphasize that the modified approach of~\cite{RescBaerByrMcC_long1997} cannot directly be applied to the two body scattering problem with a Coulomb interaction in the asymptotic configuration since the cutoff of the Coulomb potential at any $R$ distorts the asymptotic behavior of the solution at large separation of particles~\cite{FadMerk}.
In the two-body scattering problem the Coulomb potential can be implemented into the discussed approach if it is included in the free-motion Hamiltonian, while only the short-range part of the interaction is treated as the potential term.
In this case the incident wave is represented by a Coulomb wave function, which is known analytically.
This approach and its modifications are developed in papers~\cite{KruppaPRC2007,KruppaPRA2012}.

In~\cite{ourEuroLett2009, ourPRA83_2011}, we have developed a potential-splitting approach which improves the approach of~\cite{RescBaerByrMcC_long1997} in such a way that within this new formalism the ECS method can be applied to two-body systems involving the Coulomb interaction.
Instead of cutting-off the potential at a point $R$, we represent the potential as the sum $V(r) \equiv V_R(r)+V^R(r)$.
The scattering problem is first solved for the tail potential $V^R(r)$, and this solution is then used as an incident wave in the actual scattering problem.
By subtracting this incident wave from the total wave function, we obtain a function which asymptotically involves outgoing waves only, and which obeys an inhomogeneous Schr\"odinger equation.
Finally, by applying the exterior complex scaling to this equation, we obtain a boundary value problem with zero boundary conditions.
A similar idea of reformulating the original problem into a problem with zero boundary conditions was also discussed in paper~\cite{EuroPhysD2012} as well as the use of a complex basis for the Coulomb scattering problem.

In this paper we extend our splitting-potential approach to a three-particle system.
As a first step towards the solution of the overall problem, we consider in this paper the Temkin-Poet (TP) model~\cite{Temkin,Poet} of an electron scattering off a hydrogen atom and a positive helium ion.
Although the TP model is only an $S$-wave model of the three-body scattering problem, it nevertheless retains many of the essential properties and difficulties of the original scattering problem.
As such, it can be used as a test bench for various approaches to scattering calculations while keeping numerical effort moderate.
The electron scattering off ions contains an asymptotic Coulomb interaction in the incoming channel and, as it has been pointed out above for the two body case, needs a special consideration.

For the TP model of electron-hydrogen scattering, thorough studies have been performed (see e.g.~\cite{McCRescByr_ioniz1997, BartStelb_CLTP2004}) and accurate benchmark results are available~\cite{JonesStelbovics_bench,BartlettPhD}.
Conversely, detailed studies on electron-He${}^+$ calculations in the frame of the TP model are rather scarce.
However, we note results calculated with the convergent close-coupling method~\cite{MGB_Rmatrix_1995, BB_Rmatrix_1997, BC_CCC_1997} and the R-matrix method~\cite{MGB_Rmatrix_1995, BB_Rmatrix_1997, GB_Rmatrix_1997}.
For the hydrogen-like ions, results are available from the Propagating Exterior Complex Scaling (PECS) calculations~\cite{Bartlett_PhD, BarStelb_ions_2004}, while they are not reported for the TP model.
We would like to stress that this contribution is aimed at introducing a new method by which we are able to study general charged particle, three-body quantum scattering.
The present TP model study is just a first step on the way to a full angular momentum, many channel method.

The paper is structured as follows.
In Sec.~\ref{Sec_eqs}, we derive the equations underpinning the generalized splitting approach.
In the following Section~\ref{Sec_ampl}, we describe three different methods for computing scattering amplitudes and cross sections.
In Sections~\ref{Sec_eH} and  \ref{Sec_eHep}, we discuss our numerical approach and results for the TP electron-hydrogen and electron-He${}^+$ scattering, respectively.
Finally, we present our conclusions in Sec.~\ref{Sec_sum}.
Atomic units are used throughout the paper.

\section{Potential-splitting approach to the Temkin-Poet model} \label{Sec_eqs}

The TP model is a simplification of the full electron scattering problem off a hydrogen-like atom in which all angular momenta are set to zero~\cite{Temkin,Poet}.
The Hamiltonian for this model is written in terms of the electron-nucleus distances $r_1$, $r_2$ as
\begin{equation} \label{DSE_TP}
H= H_K + V(r_1,r_2),
\end{equation}
where the kinetic energy part, $H_K$, is
\begin{equation}
H_K= -\frac{1}{2} \frac{\partial^2}{\partial r^2_1}
    - \frac{1}{2} \frac{\partial^2}{\partial r^2_2}.
\end{equation}
The potential, $V(r_1,r_2)$, is the sum of the Coulomb pair-wise potentials projected on the spherically symmetric state:
\begin{equation}
V(r_1,r_2) =
-\frac{Z}{r_1}-\frac{Z}{r_2} + V_{12}(r_1,r_2),
\end{equation}
where $Z$ is the nuclear charge, and the inter-electron potential is $V_{12}(r_1,r_2) = 1/\max{\{r_1,r_2\}}$.
The wave function, $\Psi(r_1,r_2)$, as the solution to the Schr\"odinger equation
\begin{equation} \label{SE}
[H_K + V(r_1,r_2)]\Psi(r_1,r_2) = E \Psi(r_1,r_2)
\end{equation}
must both satisfy the boundary conditions $\Psi(0,r_2)=\Psi(r_1,0)=0$, and have the correct asymptotic behavior at large distances. The latter requirement will be discussed later.

As the electrons are identical fermions the proper symmetry of the wave function with respect to the permutation of the electron coordinates should be implemented.
The symmetrized wave function $\Psi^S$ is defined as
\begin{equation} \label{Psi_sym}
\Psi^S= \frac{1}{\sqrt{2}} (1+(-1)^S P_{12}) \Psi.
\end{equation}
Here $S=0,1$ represents singlet or triplet scattering, respectively.
The permutation operator $P_{12}$ interchanges the coordinates $r_1$ and $r_2$.
As the permutation operator commutes with the Hamiltonian~(\ref{DSE_TP}), the symmetrized wave function obeys the same Schr\"odinger equation $H\Psi^S=E\Psi^S$.
For the sake of clarity, our derivations will be made for the function $\Psi$, and the symmetrization will simply be done at the final stage of the formalism by applying the operator defined in Eq.~(\ref{Psi_sym}).


Let us describe the potential-splitting procedure for the three-body system within the TP model.
For definiteness we imply that electron~1 collides with the bounded complex of electron~2 and the nucleus.
Let $\chi^R(r)$ be the indicator of the domain $r\geq R$, i.e.
\begin{equation}
\chi^R(r) = \left\{
\begin{array}{ll} 0,& r < R\\ 1,& r \geq R \end{array}
\right.,
\end{equation}
and ${\chi}_R = 1-\chi^R$ be its complementary partner.
In terms of the Heaviside step function, we obtain $\chi^R(r)=\theta(r-R)$.
The potential $V^R$ is defined as:
\begin{equation}
V^R = -\frac{Z}{r_2} + \left(-\frac{Z}{r_1} + V_{12}(r_1,r_2)\right) \chi^R(r_1) ,
\end{equation}
and $V_R$ is given by
\begin{equation}
V_R = \left(-\frac{Z}{r_1} + V_{12}(r_1,r_2)\right) {\chi}_R(r_1) .
\end{equation}
The Hamiltonian can now be rewritten as the sum of three terms:
\begin{equation}
H = H_K + \left(-\frac{Z}{r_1}-\frac{Z}{r_2} + V_{12}(r_1,r_2)\right) =
H_K + V^R +V_R.
\end{equation}
The asymptotic Hamiltonian $H^R$ is defined by the expression:
\begin{equation}
H^R = H_K + V^R.
\end{equation}

First we solve the Schr\"odinger equation with the asymptotic Hamiltonian $H^R$: 
%
\begin{equation} \label{eq_psi_R}
\left[ H_K -\frac{Z}{r_2} + \left(-\frac{Z}{r_1} + V_{12}(r_1,r_2)\right) \chi^R(r_1) \right]
\Psi^R  = E \Psi^R.
\end{equation}
In order to construct the solution, the latter equation should be considered in different domains.
In the domain $r_1<R$, Eq.~(\ref{eq_psi_R}) takes the form
\begin{equation}
\left[ H_K -\frac{Z}{r_2} \right] \Psi^R = E \Psi^R.
\end{equation}
The solution regular at zero distances is found to be:
\begin{equation} \label{psiR_r1}
\Psi^R(r_1,r_2) = a^R \hat{j_0}(k_i r_1) \varphi_i(r_2),
\end{equation}
where $k_i$ is the incoming momentum, $\hat{j_0}$ is the Riccati-Bessel function, and the constant $a^R$ should be defined with the continuity condition at $r_1=R$.
The function $\varphi_i(r_2)$ is the bound state wave function for the two-body system with the Coulomb interaction:
\begin{equation}
\left( -\frac{1}{2} \frac{\partial^2}{\partial r_2^2} - \frac{Z}{r_2} \right) \varphi_i(r_2) =
\varepsilon_i\varphi_i(r_2),
\quad \varphi_i(0)=0,
\end{equation}
which is normalized such that $\int dr_2 \, \varphi_i^2(r_2) = 1$.
The energy $E$ is related to the incoming momentum as $E=\varepsilon_i+k_i^2/2$.

Now consider the domain $r_1>R$, $r_2<R$, where, Eq.~(\ref{eq_psi_R}) becomes
\begin{equation} \label{Eq_Omega2}
\left[ H_K -\frac{Z}{r_2} -\frac{Z-1}{r_1} \right] \Psi^R = E \Psi^R.
\end{equation}
The variables in the latter equation can be separated such that the solution is given by
\begin{equation} \label{psiR_Omega}
\Psi^R(r_1,r_2) = \psi^R_c(r_1,k_i) \varphi_i(r_2),
\end{equation}
and where the Coulomb scattering wave function is defined as
\begin{equation}
\psi^R_c(r_1,k_i) = e^{i\sigma_0} F_0(\eta_i,k_i r_1) + {\cal A}^R {\cal U}^+_0(\eta_i,k_i r_1).
\end{equation}
Here ${\cal A}^R$ is the scattering amplitude for the potential $V^R$.
The Sommerfeld parameter $\eta_i$ is given by $\eta_i = -(Z-1)/k_{i}$, and the s-wave Coulomb phase shift equals to ${\sigma_{0}} = \arg \Gamma(1+i\eta_{i})$~\cite{Abramowitz}.
The outgoing Coulomb wave
\[
{\cal U}^+_0(\eta_i,k_i r_1) = e^{- i\sigma_0}
\left[G_{0}(\eta_{i},k_i r_1) + iF_{0}(\eta_{i},k_i r_1)\right]
\]
is defined in terms of the regular (irregular) Coulomb wave $F_{0}\,(G_{0})$.

In the domain $r_1>R$, $r_2>R$, the relationship between $r_1$ and $r_2$ must be taken into consideration in order to construct the solution.
When $r_1> r_2$, Eq.~(\ref{eq_psi_R}) coincides with Eq.~(\ref{Eq_Omega2}), and the solution can be expressed in the form  (\ref{psiR_Omega}), i.e. $\psi^R_c(r_1,k_i) \varphi_i(r_2)$.
In the opposite case of $r_2> r_1 >R$, the asymptotic Hamiltonian $H^R$ changes to
\begin{eqnarray}
H^R & = & H_K -\frac{Z}{r_2} + \left( - \frac{Z}{r_1} + \frac{1}{r_2} \right) \nonumber \\
		& = & \left( H_K -\frac{Z}{r_2} - \frac{Z-1}{r_1} \right) + \left( \frac{1}{r_2} - \frac{1}{r_1} \right),
\end{eqnarray}
and, therefore,
\begin{eqnarray} \label{psi phi}
\left(H^R - E\right) \psi^R_c(r_1,k_i) \varphi_i(r_2)   \nonumber \\
   = \left( \frac{1}{r_2} - \frac{1}{r_1} \right) \psi^R_c(r_1,k_i) \varphi_i(r_2).
\end{eqnarray}
Collecting the results obtained so far, the solution to the Schr\"odinger equation~(\ref{eq_psi_R}) can be represented as the sum of two terms
\begin{equation} \label{Psi^R}
\Psi^R(r_1,r_2)=\psi^R_c(r_1,k_i) \varphi_i(r_2) +U^R(r_1,r_2).
\end{equation}
Here the residual term $U^R$ obeys the inhomogeneous equation
\begin{eqnarray} \label{UR}
\left(H^R - E\right) U^R(r_1,r_2)   \nonumber \\
=  \left( \frac{1}{r_2} - \frac{1}{r_1} \right) \psi^R_c(r_1,k_i) \varphi_i(r_2)\theta(r_2-r_1)\chi^R(r_1),
\end{eqnarray}
where the right hand side is restricted to the region where $r_1>R$ and $r_2>r_1$. 
The combination of functions $\varphi_i(r_2)\theta(r_2-r_1)$ makes the right hand side of Eq.~(\ref{UR}) fast decreasing in both coordinates $r_1$ and $r_2$ for any $R$. 
This is the property of the equation which is required for the application of the complex scaling method, hence the solution $U^R$ can be obtained by the same method as we use to construct the solution of our final driven Schr\"odinger equation (see below). 
If $R$ is large then due to the presence of $\chi^R(r_1)$, the right hand side of Eq.~(\ref{UR}) is not zero if $r_2>R$ and the wave function $\varphi_i(r_2)$ is exponentially small whereas the rest is bound.
Under these conditions, the solution to Eq.~(\ref{UR}) is exponentially small and therefore is negligible with respect to the first term in the right hand side of Eq.~(\ref{Psi^R}) which does not decrease for large $R$. 
Our numerical calculations support this statement since we have not found any noticeable influence of the function $U^R$ on our results at large $R$.
However, for arbitrary value of $R$ one should keep $U^R$ in the formalism for the completeness.


It is left to determine the constants $a^R$ and ${\cal A}^R$ for the function $\psi^R_c(r_1,k_i)$, where $\psi^R_c(r_1,k_i)$ is the solution to the equation
\begin{equation}
\left(-\frac{d^2}{d r_1^2}  -\frac{(Z-1)\chi^R(r_1)}{r_1}-k_i^2\right)\psi^R_c(r_1,k_i)=0.
\end{equation}
Here, the solution is given by
\begin{equation}
\psi_c^R(r_1,k_i) =
\left\{
\begin{array}{ll}
a^R \hat{j_0}(k_i r_1), & r_1 < R. \\
\left[ e^{i\sigma_0} F_0(\eta_i,k_i r_1) \right. \\
\left. + {\cal A}^R {\cal U}^+_0(\eta_i,k_i r_1)\right] ,
& r_1 \geq R.
\end{array}
\right.
\end{equation}
where the values of $a^R$ and ${\cal A}^R$ are explicitly calculated from the matching conditions at the point $r_1 = R$ (see~\cite{ourPRA83_2011}).
We will see that only $a^R$ contributes to the inhomogeneous term of the driven Schr\"odinger equation which needs to be solved.
The explicit form for this constant is given by:
\begin{equation} \label{a_R}
a^R = k_i [{\cal U}^+_0(\eta_i,k_i R) \hat{j_0}'(k_i R) -
{\cal U}^{+'}_0(\eta_i,k_i R) \hat{j_0}(k_i R)]^{-1}.
\end{equation}
%

Now we have all what is needed to rewrite the Schr\"odinger equation~(\ref{DSE_TP}) in the driven form which is the main tool of the potential-splitting approach.
The total wave function $\Psi$ is represented as the sum of two terms
 \begin{equation} \label{Psi split}
 \Psi=\Psi^R+\Phi.
 \end{equation}
The Schr\"odinger equation now reads
\begin{equation}
H\Psi^R+H\Phi=E(\Psi^R+\Phi).
\end{equation}
Using the Hamiltonian splitting $H=H^R+V_R$ and the fact that $(H^R-E)\Psi^R=0$, we obtain the desired driven equation
\begin{equation}
(H-E)\Phi=-V_R \Psi^R.
\end{equation}
In its explicit form, the latter equation is written as
\begin{eqnarray} \label{driven_eq}
\left[H_K + \left(-\frac{Z}{r_1}-\frac{Z}{r_2} + V_{12}(r_1,r_2)\right) -E\right]
\Phi(r_1,r_2) \nonumber \\
= -\left[-\frac{Z}{r_1} + V_{12}(r_1,r_2)\right] {\chi}_R(r_1)
\Psi^R(r_1,r_2).
\end{eqnarray}

As mentioned above, the wave function for the system should be properly symmetrized with respect to the permutation of electrons.
Applying operator defined in Eq.~(\ref{Psi_sym}) to Eq.~(\ref{driven_eq}), we get for the symmetrized function $\Phi^S$,
\begin{equation}
\Phi^S= \frac{1}{\sqrt{2}} (1+(-1)^S P_{12}) \Phi,
\end{equation}
the final form of desired driven Schr\"odinger equation:
\begin{eqnarray} \label{driven_eq_sym}
\left[H_K + \left(-\frac{Z}{r_1}-\frac{Z}{r_2} + V_{12}(r_1,r_2)\right) -E\right]
\Phi^S(r_1,r_2) \nonumber \\
 = - \frac{1}{\sqrt{2}} (1+(-1)^S P_{12}) \nonumber \\
\times \left(-\frac{Z}{r_1} + V_{12}(r_1,r_2)\right) {\chi}_R(r_1)
 \Psi^R(r_1,r_2) .
\end{eqnarray}
It is worth noting that the right hand side of Eq.~(\ref{driven_eq_sym}) decreases asymptotically in both coordinates $r_1$ and $r_2$ since ${\chi}_R(r_1)$ is zero for $r_1>R$, and the function $\Psi^R(r_1,r_2)$ decreases exponentially in $r_2$ due to the bound state wave function $\varphi_i(r_2)$  and properties of $U^R(r_1,r_2)$ for large $r_2$.
These properties of Eq.~(\ref{driven_eq_sym}) are decisive for applicability of the ECS method.
Using this method, the coordinates $r_1$ and $r_2$ are rotated at a sufficiently large radius, $Q$, into the complex plain by a fixed angle $\theta$.
The transformation is chosen to be
\begin{equation}
z(r) = \left\{
\begin{array}{ll}
r, & r<Q \\
Q+f(r-Q) e^{i\theta}, & r \geq Q,
\end{array}
\right.
\end{equation}
where $f(t)$ is constructed such that both $z(r)$ and $z'(r)$ are continuous, see e.g.~\cite{ElaYar_helium1998}.
The right hand side of Eq.~(\ref{driven_eq_sym}) decreases exponentially for large $R$, $r_1 \geq R$ and $r_2 \geq R$, so that the exterior complex scaling transformation can be directly applied with $Q\ge R$.
We have found no reason to have distinct values for the splitting $R$ and exterior rotation $Q$ points although this is not a requirement.

\section{Calculation of amplitudes and cross sections} \label{Sec_ampl}

On solving Eq.~(\ref{driven_eq_sym}) with the ECS, we obtain the wave function $\Phi(r_1,r_2)$ in the region $r_1, r_2 \leq R$.
The next step is to calculate the amplitudes and cross sections corresponding to the various scattering processes occurring in the system.
There exists extensive literature which discusses the different methods available for calculating both elastic and breakup (ionization)~\cite{JonesStelbovics_bench, Bartlett_PhD, McCResc_breakup2000, Kadyrov_2004} amplitudes.

All of these methods are based on the asymptotic form of the wave function at large distances \cite{FadMerk,Rudge}:
\begin{eqnarray} \label{wf_asympt}
\Phi^S(r_1,r_2) \sim \sum_j f_{ji}^S(k_j)
\frac{1}{\sqrt{2}} (1+(-1)^S P_{12}) \nonumber \\
\times {\cal U}^+_0(\eta_j,k_j r_1) \varphi_j(r_2)
+F^S(\alpha) \frac{e^{iK\rho+iQ_0(\alpha,\rho)}}{ (K\rho)^{5/2}}.
\end{eqnarray}
Here the scattering amplitudes $f_{ji}^S$ correspond to the elastic and inelastic channels moving the atom to the discrete state $\varphi_j(r)$ from the initial state $\varphi_i(r)$.
The second term on the right hand side represents the breakup with the amplitude $F^S(\alpha)$, and it is expressed in terms of the hyperradius $\rho=(r_1^2+r_2^2)^{1/2}$ and the angle $\alpha=\arctan(r_2/r_1)$.
The logarithmic phase factor is defined as $Q_0(\alpha,\rho)= -\frac{\rho}{K} V(r_1,r_2)\ln{(2K\rho)}$.
The break-up momentum $K$ is given in terms of the  energy by the formula $K^2=2 E$.

The amplitude $f_{ji}^S$ can be calculated from the projection of the wave function on the two-body bound states $\varphi_j(r_2)$.
It is defined as a limit of the projected wave function at infinity:
\begin{equation} \label{ampl_proj}
f_{ji}^S(k_j) = \lim_{r_1 \to \infty} \sqrt{2} \left({\cal U}^+_0(\eta_j,k_j r_1)\right)^{-1}
\!\! \int\limits_0^\infty \! \! d r_2 \,  \varphi_j(r_2) \Phi^S(r_1,r_2).
\end{equation}
The symmetrized term is neglected as the discrete state wave function decreases exponentially with $r_1$.
By using the ECS approach, we calculate the wave function in the region $r_1, r_2 \leq R$.
This means that we should restrict the integration in Eq.~(\ref{ampl_proj}) to the interval $r_2 \in [0,R]$ and consider the value of the right hand side at some $r_1 \leq R$.
The calculated amplitude converges to the exact value when $R\to \infty$.
The elastic and inelastic cross sections are given in terms of the amplitudes by
\begin{equation}
\sigma_{ji}^S(k_j) = 4\pi \frac{k_j}{k_i} |f_{ji}^S(k_j)|^2.
\end{equation}
The cross sections can also be computed using the representation based on the projected optical theorem~\cite{McCRescByr_ioniz1997}
\begin{equation} \label{ampl_opt_theor}
\sigma_a^S = -\frac{8\pi}{k_i^2}
\Im\mbox{m} \int\limits_0^R \! \left[P_a \Phi^S(R,r_2)\right]^*
\frac{\partial}{\partial r_1} \left.\left[P_a \Phi^S(r_1,r_2)\right]\right|_{r_1=R}\, dr_2.
\end{equation}
Here the subscript $a$ denotes either the total, ionization or excitation cross section, and $P_a$ represents the corresponding projection operator~\cite{McCRescByr_ioniz1997}.

An alternative approach to obtain the required amplitudes is given by applying the surface integral representation~\cite{BartStelb_CLTP2004}:
\begin{eqnarray} \label{ampl_surface}
f_{ji}^S(k_j) = \sqrt{2} \int\limits_0^R \! d r_2 r_1^2 \varphi_j(r_2)
\left[
  \frac{\Phi^S(r_1,r_2)}{r_1}  \frac{\partial}{\partial r_1} j_0(k_j r_1) \right. \nonumber  \\
  \left. - j_0(k_j r_1) \frac{\partial}{\partial r_1} \frac{\Phi^S(r_1,r_2)}{r_1}
\right].
\end{eqnarray}
Here $j_0(r)=\hat{j_0}(r)/r$ is the spherical Bessel function, and where, on the right hand side, $r_1$ should be set as $r_1=R$.

The calculation of the differential ionization cross section is a special case, and has been extensively discussed in the literature (see, e.g.~\cite{McCRescByr_ioniz1997, Bartlett_PhD, Kadyrov_2004, BaerRescMcC_ampl_2001} and references therein).
In this report, however, we focus mainly on the calculation of the wave function, and we consider the differential ionization cross section outside the scope of the paper.
The detailed study of the ionization cross section will be done elsewhere.

\section{Electron-hydrogen scattering} \label{Sec_eH}

For the case of electron-hydrogen scattering, $Z=1$ and $\eta_i=0$.
Hence Eq.~(\ref{a_R}) gives $a^R=1$, and Eq.~(\ref{driven_eq_sym}) coincides with the equations used in the papers~\cite{McCRescByr_ioniz1997, BartStelb_CLTP2004, JonesStelbovics_bench} to study the electron-hydrogen TP model.
Here, we use this model in order to compare the results of our calculations with the above mentioned results and to check the accuracy and stability of our numerical approach.

The numerical approach employed here is based on the finite element method (FEM) combined with the ECS method, and has been previously used for calculating resonances in three-body quantum systems~\cite{ElaYar_helium1998, ElaLevYar_NeICl2009}.
From the numerical point of view, the solution of the scattering problem is straightforward compared to the calculation of resonances as we only need to solve the system of linear algebraic equations instead of a generalized eigenvalue problem.
This issue is briefly discussed in paper~\cite{ourFewBody2009}.
Conversely, the physical space (and therefore the matrix dimensions) of the problem is larger for the scattering calculations.

In the calculations, we use a rectangular grid formed by the same one-dimensional grid in both coordinates $r_1$ and $r_2$.
For each coordinate, we use five finite elements at short distance [0-4]~a.u., and four elements of total length 40~a.u. for the discretization beyond the rotation point $R$.
The intermediate region is divided into elements with a length 3~a.u.
With this grid the number of the finite elements and the size of the FEM matrix depends on the radius $R$.
For most calculations, the polynomial degree in the FEM was chosen to be seven.
The maximum size of the matrix in our calculations is 166464 with a sparseness about $2 \cdot 10^{-4}$.
The value of the complex rotation angle $\theta$ has little influence on the results and was chosen to be $\theta=45^o$.

\begin{figure}[t]
\centering
\includegraphics[width=0.48\textwidth]{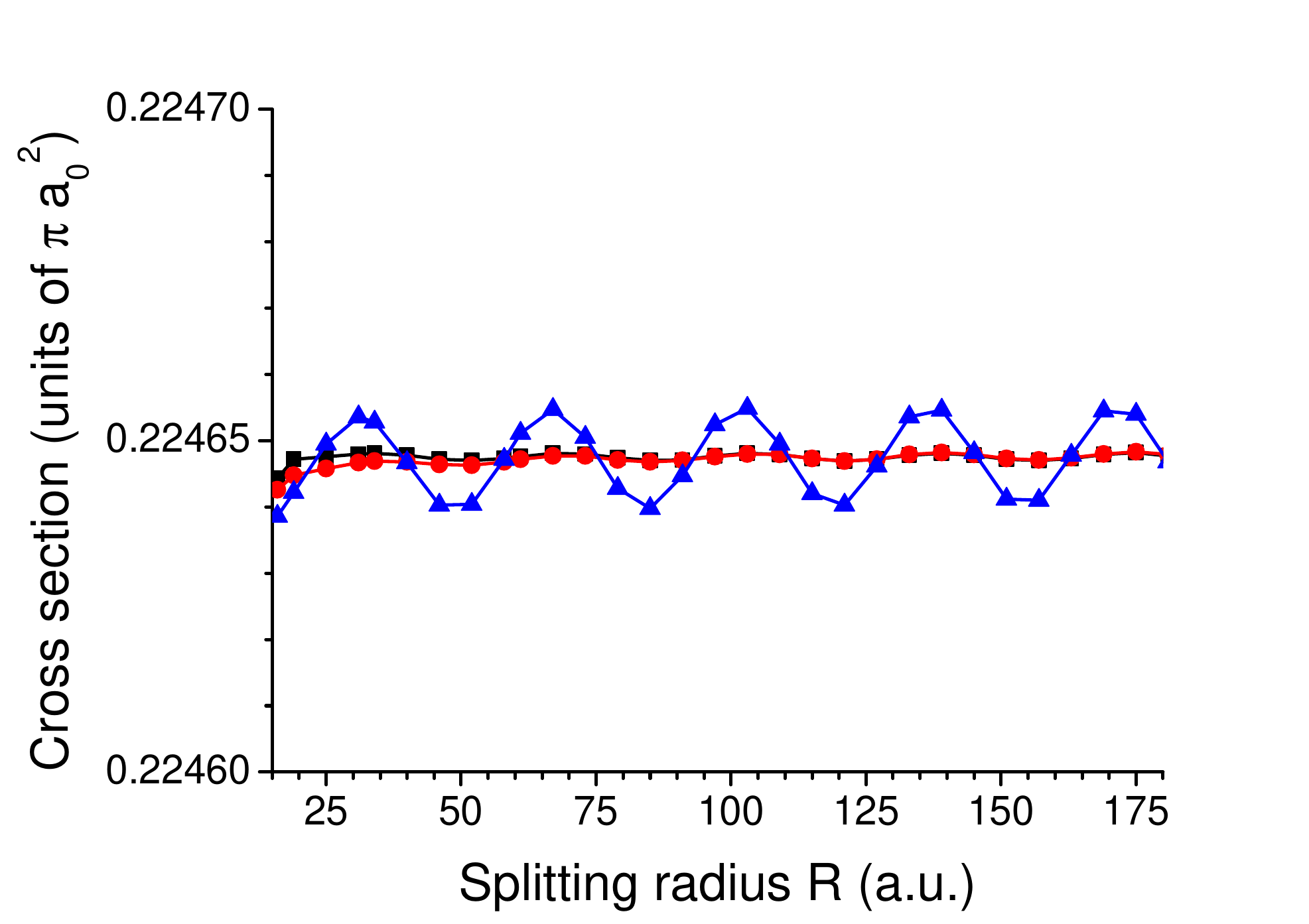}
 \caption{(Color online) The TP singlet $1s$ scattering cross section at energy $E=17.6$~eV with spin weighting, as a function of the radius $R$.
The squares, circles, and triangles are the results calculated with the integration formula, Eq. ~(\ref{ampl_proj}), the projected optical theorem, Eq. ~(\ref{ampl_opt_theor}), and the surface integral, Eq.~(\ref{ampl_surface}), respectively.}
 \label{figTP-1}
\end{figure}
\begin{figure}[t]
\centering
\includegraphics[width=0.48\textwidth]{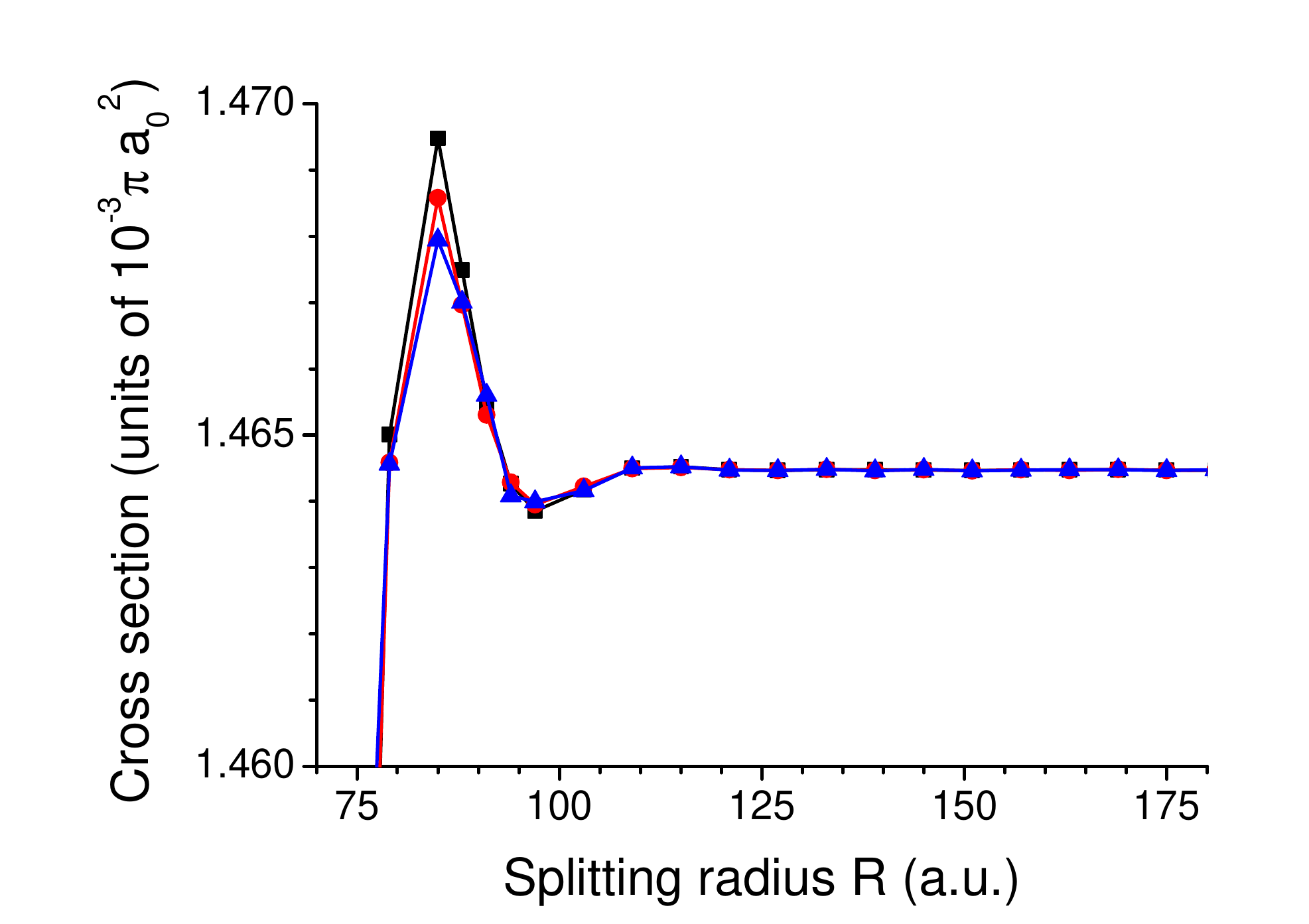}
 \caption{(Color online) The TP singlet $5s$ scattering cross section. The symbols are the same as in Fig.~\ref{figTP-1}. }
 \label{figTP-5}
\end{figure}

Figs.~\ref{figTP-1} and \ref{figTP-5} show the convergence for the singlet $1s$ and $5s$ scattering cross sections as a function of the splitting (and rotation) radius $R$.
The results obtained with all representations~(\ref{ampl_proj},\ref{ampl_opt_theor},\ref{ampl_surface}) give comparable accuracy for the different output channels.
The surface integral representation, Eq.~(\ref{ampl_surface}), shows some oscillations for the $1s$ state while it is more stable for calculations of the excited states.
We suggest the reason for this behavior is that the two-body wave function $\varphi_i(r_2)$ does not vanish at small $r_2$, resulting in interference with the ionization term at large separations.
Similar oscillations are also observed from the two other methods, although their amplitudes are considerably smaller.
The slower convergence of the $5s$ scattering cross section with respect to $R$, as displayed in Fig.~\ref{figTP-5}, is not surprising given that the spatial extension of the $n^{th}$ two-body states grows rapidly with $n$.

We have compared our elastic and inelastic  cross sections calculated with Eq.~(\ref{ampl_proj}) for $R=181$~a.u. with the results reported in paper~\cite{JonesStelbovics_bench} and found them to be in very good agreement.
For both the triplet and singlet excitation cross sections at energies 17.6~eV -- 54.4~eV, the relative discrepancy of our results and results of paper~\cite{JonesStelbovics_bench} is less than 0.1\%.
The agreement cannot be expected much better due to the final number of quoted digits in the results.
At higher energy, $E=150$~eV, the results differ up to 1\%.
The reason for the larger differences is a faster oscillation of the wave function, and so a higher computational accuracy is needed.
This accuracy can be achieved, e.g., by increasing the polynomial order in the FEM.
For instance, an increase the polynomial order by one decreases the maximum relative discrepancy to 0.2\%.

In order to check accuracy of our cross sections at lower energies, we have compared them to the results presented in paper~\cite{BartlettPhD} where the data for energies 2.7~eV -- 54.4~eV are available.
The discrepancy of the results at all given energies and for all states is less than 0.2\%.
The only exception is the energy of the ionization threshold itself, $E=13.6$~eV, where the results differ by as much as 3\% for higher excited states.
The reason is that at small relative energy, a large space is necessary for the scattering wave function to establish its asymptotic behavior.
In this case, the radius $R$ has to be increased to get even better accuracy.
Also we would like to mention that the convergence of our results as $R\to \infty$ does not shows any essential oscillatory character and therefore for final result we do not need any averaging  unlike it is done in the approach of paper~\cite{BartlettPhD}.

\begin{figure}[t]
\centering
\includegraphics[width=0.48\textwidth]{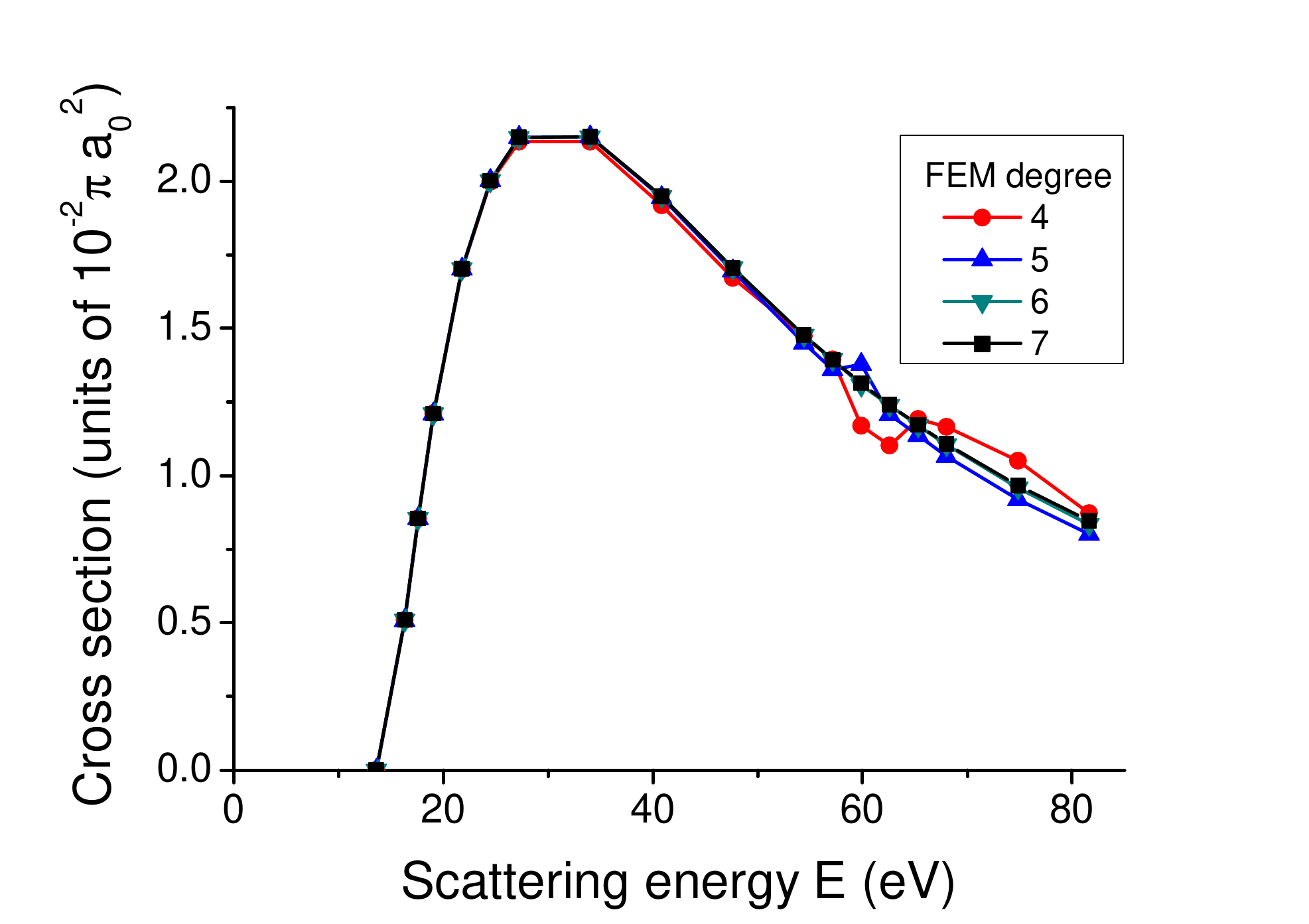}
 \caption{(Color online) The singlet total ionization cross section as a function of the energy for different orders of the FEM polynomial.
 The radius $R=181$~a.u.}
 \label{figTP-pol}
\end{figure}

The dependence of the accuracy on the polynomial order in the FEM is illustrated with the calculation of the total ionization cross section~(\ref{ampl_opt_theor}).
The projection operator $P_{\rm ion}$ is defined in terms of the projections on the two-body states $P_i$~\cite{McCRescByr_ioniz1997} as follows
\begin{equation} \label{P_ion}
P_{\rm ion} = I-\sum_{i=1}^{M} P_i.
\end{equation}
Fig.~\ref{figTP-pol} shows the calculated singlet total ionization cross sections for $M=20$.
At higher energies, the scattering wave function oscillates more rapidly so higher polynomial order is required to achieve better accuracy.
The difference between the values calculated for two subsequent polynomial degrees can be used as a crude estimation of the accuracy.
Furthermore, more accurate estimations can be constructed based on the known error behavior in the FEM~\cite{ElaYar_helium1998}.
In the vicinity of the ionization threshold the cross section approaches zero.
However, its accurate calculation may be necessary, e.g., for reproducing threshold laws~\cite{BartStelb_CLTP2004}.
For such calculations, the radius $R$ has to be increased to get better accuracy as discussed above.

\begin{figure}[t]
\centering
\includegraphics[width=0.48\textwidth]{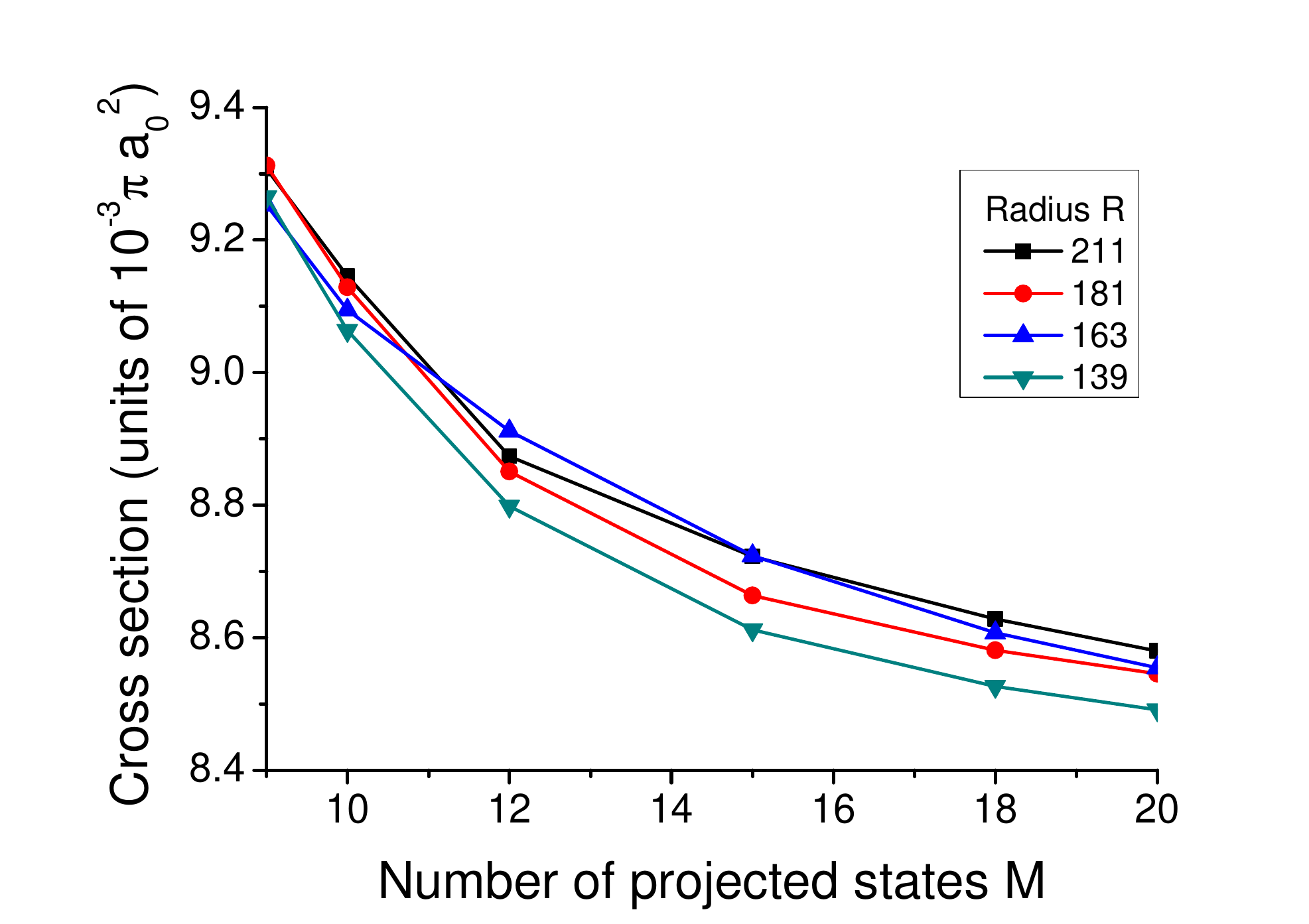}
 \caption{(Color online) The singlet total ionization cross section as a function of the number of projected states $M$ for different values of the splitting radius $R$.
The energy E=17.6~eV.}
 \label{figTP-proj}
\end{figure}

Fig.~\ref{figTP-proj} displays the same cross section as a function of the number of projected states $M$ in Eq.~(\ref{P_ion}) for different values of the splitting radius $R$.
The cross section decreases with the increase in the number of projected states, and increases with increasing splitting radius.

\section{Electron-He${}^+$ scattering} \label{Sec_eHep}

\begin{figure}[t]
\centering
\includegraphics[width=0.48\textwidth]{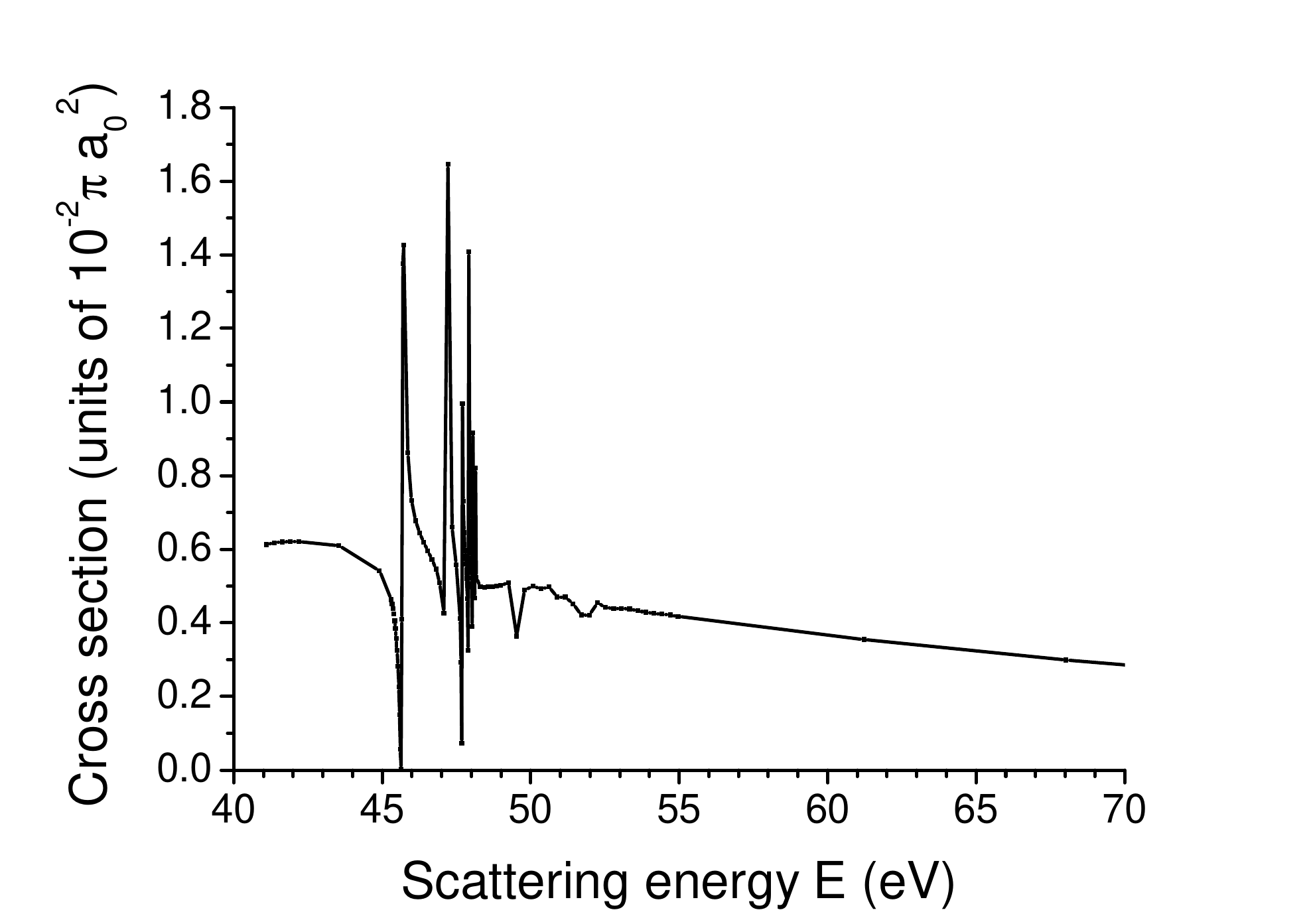}
 \caption{The singlet (spin weight included) 2$S$ excitation cross section for the TP model of $e$-He${}^+$ scattering as a function of the incident electron energy. }
 \label{fig-eHe-S1s2s}
\end{figure}

The main goal of the present contribution is to describe scattering systems with Coulomb asymptotic behavior in the incoming and outgoing channels.
There exist a few studies of electron-He$^+$ scattering in the literature.
We would like to mention results calculated with the the convergent close-coupling method~\cite{MGB_Rmatrix_1995, BB_Rmatrix_1997, BC_CCC_1997}, the R-matrix method~\cite{MGB_Rmatrix_1995, BB_Rmatrix_1997, GB_Rmatrix_1997} which all are applied to the TP model.
Using Eq.~(\ref{driven_eq_sym}) derived here, we have calculated the excitation  cross sections for the electron-He${}^+$ scattering.
Fig.~\ref{fig-eHe-S1s2s} displays our results for the 1$s$-2$s$ singlet excitation cross section.
A number of resonances are observed in this system.
Some peaks in the cross section may not be visible in the figure due to the finite steps in the energies used in the calculations.
The resonance peaks are well pronounced and accumulate to the thresholds from below.
A proper comparison of the resonant structures in Fig.~\ref{fig-eHe-S1s2s} and resonance study should not only include the resonances energies but also the influence of these possible resonances on the cross section as demonstrated in the two-body study in paper~\cite{Kseniahh3}.
The generalization of that technique seems to be relatively straightforward but remains to be done in detail.
Comparing the results in papers~\cite{MGB_Rmatrix_1995, BB_Rmatrix_1997, BC_CCC_1997,GB_Rmatrix_1997} to our results we find that our approach demonstrates both good stability and a high accuracy for the scattering problem.
In contrast to the close-coupling and $R$-matrix approaches~\cite{BB_Rmatrix_1997}, our calculations do not suffer from artificial noise and oscillations.

Our approach has some similarity with the PECS calculations reported in~\cite{Bartlett_PhD, BarStelb_ions_2004}.
As it was mentioned above PECS results are not reported for the TP model. 
It is why we adopted the PECS formalism to the TP model for comparing its results with ours.
It is important to note that Eq.~(\ref{driven_eq_sym}) differs from the equations derived in the frame of the PECS approach.
Indeed, if we restrict the driving term of the PECS equation~\cite{BarStelb_ions_2004} for the TP model (i.e. set the total and electron angular momenta to zero), this term can be written in the current notations as
\begin{eqnarray} \label{PECS_rhs}
  - \frac{1}{\sqrt{2}} (1+(-1)^S P_{12}) \left(\frac{1}{r_2} - \frac{1}{r_1} \right)
  \theta(r_2-r_1) \nonumber \\
 \times e^{-i\sigma_0} F_0(\eta_i,k_i r_1) \varphi_i(r_2).
\end{eqnarray}
The left hand side of the PECS equation is identical to that of Eq.~(\ref{driven_eq_sym}).
In fact, the driven equation in paper~\cite{BarStelb_ions_2004} is constructed in such a way that the analog of the function $\Psi^R$ is chosen as the product of the Coulomb wave corresponding to the electron scattering off the charge $Z-1$, and the two-body bound state of the electron and the nucleus of charge $Z$.
This choice corresponds to an asymptotic Hamiltonian with the Coulomb interaction between the electron and an infinitely heavy particle with a charge $Z-1$.
The PECS equation cannot be directly derived from our equation only with an appropriate choice of $R$ in Eq.~(\ref{driven_eq_sym}).

\begin{figure}[t]
\centering
\includegraphics[width=0.48\textwidth]{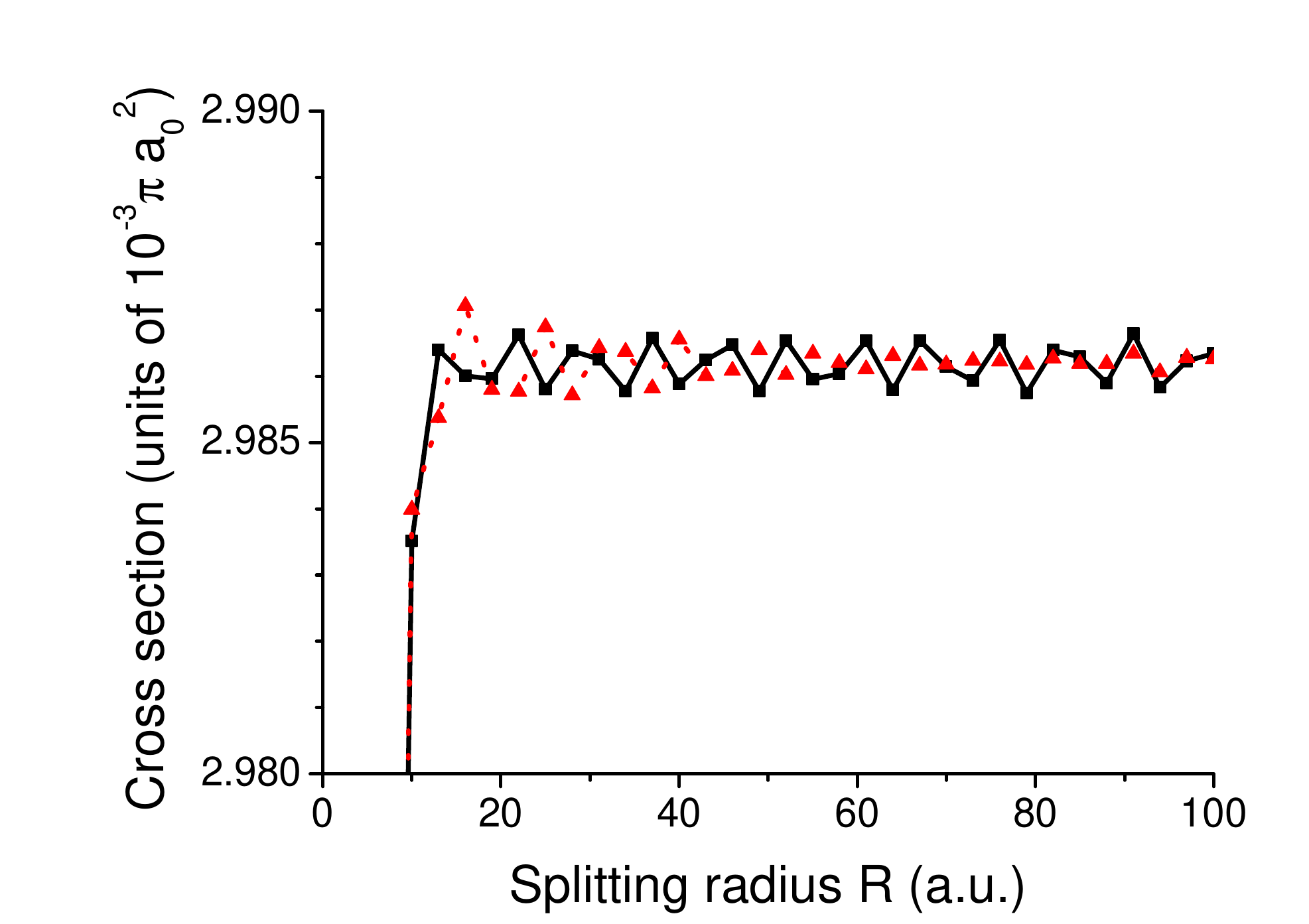}
 \caption{(Color online) The singlet (spin weight included) 2$S$ excitation cross section for the TP model of $e$-He${}^+$ scattering at $E$=2.5~a.u. as a function of the splitting radius $R$.
 The results obtained with Eq.~(\ref{driven_eq_sym}) and Eq.~(\ref{PECS_rhs}) are displayed as triangles and squares, respectively.}
 \label{fig-eHe-diff}
\end{figure}

Fig.~\ref{fig-eHe-diff} shows an example of the 1$s$-2$s$ singlet excitation cross section as a function of the splitting radius $R$ for the PECS equation~(\ref{PECS_rhs}) and Eq.~(\ref{driven_eq_sym}).
These results are computed with the identical numerical grids described in the previous section.
While the results for any specific radius are slightly different, they converge to the same value and their accuracies are comparable.
Hence both Eq.~(\ref{driven_eq_sym}) and Eq.~(\ref{PECS_rhs}) can be successfully used for scattering calculations.

\section{Conclusions} \label{Sec_sum}

The potential-splitting approach is successfully applied to model three body systems with both neutral and Coulomb asymptotic behavior in the incoming and outgoing channels.
We compare our elastic and inelastic cross sections for the electron-hydrogen scattering with the most accurate data available~\cite{JonesStelbovics_bench,BartlettPhD} and find a very good agreement.
We would further stress the advantage of the present approach in the sense that observed, resonant peaks in the cross sections can in principle be verified to be caused by resonances and  be computed and characterised with the approach described in our previous studies~\cite{ElaYar_helium1998, ElaLevYar_NeICl2009}.

While the propagating exterior complex scaling (PECS) equations~\cite{Bartlett_PhD, BarStelb_ions_2004} cannot be directly deduced from Eq.~(\ref{driven_eq_sym}), they are based on the construction of the solution with an asymptotic Coulomb interaction.
In fact, the Coulomb interaction in the PECS is present in the entire space.
On the contrary, the potential-splitting approach can be used for studying systems with a general interaction featuring an asymptotic Coulomb tail, e.g. molecular ion systems described with point-wise computed \textit{ab initio} potential energy surfaces.
We have also derived the function $U^R(r_1,r_2)$~(\ref{UR}) which shows the exact effect of the potential cut-off and can be used as the estimation for inaccuracy caused by the potential cut-off at large distances in the exterior complex scaling approach.
The developed approach can also be generalized for higher partial waves using the results for the three-dimensional sharp screening~\cite{our_JPA2010, FewBody2014}.
This work as well as development of towards many channel quantum scattering of charged diatomic molecules against atomic ions are in progress.

\acknowledgments

The work of SLY and EY was supported in part by Russian Foundation for Basic Research with grant No.~14-02-00326 and by St Petersburg State University within the project No.~11.38.263.2014.
SLY and EY are grateful for support made possible under the bilateral agreement between St Petersburg State University and Stockholm University.
The work of EY was also financed by a grant from the Stockholm University Academic Initiative (2011) by which Stockholm University has supported visits from its Sister Universities.
Finally we also acknowledge support from the Swedish Research Council and the Carl Trygger foundation.


\end{document}